\documentclass[showpacs, twocolumn]{revtex4-1}

\usepackage{graphicx}
\usepackage{amsmath}
\begin{document}

\title{Two-dimensional dipolar bosons with weak disorder}

\author{Abdel\^{a}ali Boudjem\^{a}a}

\affiliation{Department of Physics, Faculty of Sciences, Hassiba Benbouali University of Chlef P.O. Box 151, 02000, Ouled Fares, Chlef, Algeria}

\email {a.boudjemaa@univ-chlef.dz}

\begin{abstract}

We consider two-dimensional dipolar bosonic gas with dipoles oriented perpendicularly to the plane in a weak random potential. 
We investigate analytically and numerically the condensate depletion, the one-body density-matrix, the ground state energy, the sound velocity and the superfluid fraction.
Concentrating on the regime where a rotonlike excitation spectrum forms, our results show that the superfluidity disappears 
below a critical value of disorder strength yielding the transition to a non-trivial quantum regime.
\end{abstract}


\maketitle

%\section{Introduction}

Recently, ultracold dipolar gases in two-dimentional (2D) geometry have been the subject of intense experimental  and theoretical
investigations \cite{Baranov, Pupillo2012, Osp, Wu, Tak}. 
What renders such systems particularly intriguing is the presence of the low-lying roton minimum in
the excitation spectrum \cite {gora, boudjGora} and the possibility of the crystallization of solid bubble
into a lattice superstructure, resulting in a global supersolid phase \cite {prok, boudjGora, petGora}.
However, such supersolids require a dense regime with several particles within the interaction range, which can be
difficult to achieve. The same holds for supersolids discussed for 2D dipolar Bose gases \cite{Kurb} near the gas-solid
phase transition \cite{Buch, Ast}. It was found also that this state appears  in B-t-J model of two-component bosons as a result of the long-range DDI \cite{Yos}.

A more complicated situation arises if these phenomena are studied in a random environment. 
Bose gases in the random medium attracted a great deal of interest because it connects two central ideas of the condensed matter theory: 
the Bose-Einstein condensation (BEC) and localization. This latter occurs both for fermions and bosons \cite {Fisher} and results in the existence of an insulating
phase called “Bose glass” \cite {Fisher}.
The existence of a phase characterized by simultaneous glassiness and superfluidity and thus constitutes a glassy
counterpart to the supersolid phase \cite {And, Ches, Legget}, was first observed in numerical Quantum Monte Carlo simulations of solid ${}^4$He samples 
by Boninsegni et\textit {al}\cite {Mass}. This phase which is featured (at the same time) by superfluidity and a metastable amorphous structure called "superglass".
%Other important issue of the SG phase is that it has a low-temperature plateau in the superfluid fraction \cite {Mass}.
Superglass may also have realizations in interacting bosons at very low temperatures and high density \cite {prok, Biroli}.
In this context, Zamponi et\textit {al} \cite{Zamp} have shown that quantum fluctuations can stabilize the superglass phase in a self-disordered environment induced by geometrical frustraion.
%Superglass may also have realizations in other areas such as: superconductivity "lattice SG" \cite {Biroli, Yin} and spin systems \cite {Seng}. 
%and the transition-metal compounds \cite {Brink}. Moreover, SG can occur also in classical viscous systems with quenched disorder\cite{Mez}. 

On the other hand,  many interesting works on BEC and superfluidity (See, e.g., \cite {Huang,Gior,Lopa}) have
been reported for disordered cold atomic gases with pure contact interaction in continuum, the study on dipolar boson systems is still inadequate.
In the present paper, for the first time to our knowledge,  we investigate the properties of a  quasi-2D homogeneous dipolar Bose gas in a weak random potential 
with delta correlated disorder. 
Our study is based on the Bogoliubov approach, this method which marked an important step towards a quantitative description of dirty dipolar Bose systems \cite{AxelM, Boudj}, 
allows for accurate determination of the condensed depletion, one-body density matrix, ground state energy, sound velocity
and superfluid fraction. It is found that the presence of a disordered potential in the regime where the roton develops in the excitations spectrum 
strongly enhances fluctuations and thermodynamics quantities.
We demonstrate that both BEC and superfluidity are depressed due to the competition between disorder and DDI yielding, the transition to an unusual quantum phase.

%\section{ The Model}

We consider a dilute Bose-condensed gas of dipolar bosons in an external random potentials. These particles can be confined to quasi-2D, by means of an external harmonic potential
in the direction perpendicular to the motion (pancake geometry) and all dipoles are aligned perpendicularly to the plane of their translational
motion, by means of a strong electric (or magnetic) field. 
In this quasi-2D geometry, at large interparticle separations $r$ the interaction potential is  
$V(r) = d^2/ r^3=\hbar^2r_*/mr^3$, with $d$ being the dipole moment, $m$ the particle mass, and $r_*=md^2/\hbar^2$ the characteristic dipole-dipole distance.
The disorder potential is described by vanishing ensemble averages $\langle U(\vec r)\rangle=0$
and a finite correlation of the form $\langle U(\vec r) U(\vec r')\rangle=R (\vec r,\vec r')$.

%\begin{figure}[htb1]
%\includegraphics[scale=0.5]{ddi.eps}
%\caption (Color online) {Dipolar Bose-Einstein condensate tightly confined in one direction.}
%\end{figure}

In the ultracold limit where the particle momenta satisfy the inequality $kr_*\ll1$, the scattering amplitude is given by (see e.g. \cite{boudjGora})
\begin{equation}\label{ampl} 
 f({\vec k},{\vec k'})=g(1-C\vert \vec k-\vec k'\vert),
\end{equation}
where the 2D short-range coupling constant is $g=g_{3D}/\sqrt{2}l_0$ with $l_0=\sqrt{\hbar/m \omega}$, $\omega$ is the confinement frequency
and $C =2\pi \hbar^2r_*/mg=2\pi d^2/g$.
Employing this result in the secondly quantized Hamiltonian \cite{boudjGora}, we obtain

\begin{align}\label{he3}
\hat H\!\!=&\!\!\sum_{\vec k}\!E_k\hat a^\dagger_{\vec k}\hat a_{\vec k}\! +\!\frac{1}{S}\!\!\sum_{\vec k,\vec p} \! U_{\vec k\!-\!\vec p} \hat a^\dagger_{\vec k} \hat a_{\vec p} \\ \nonumber
&+\!\frac{g}{2S}\!\!\sum_{\vec k,\vec q,\vec p}\!\!
(1\!\!-\!C\vert \vec q\!-\!\vec p\vert)\hat a^\dagger_{\vec k\!+\!\vec q} \hat a^\dagger_{\vec k\!-\!\vec q}\hat a_{\vec k\!+\!\vec p}\hat a_{\vec k\!-\!\vec p} ,
\end{align}
where $S$ is the surface area, $E_k=\hbar^2k^2/2m$, and $\hat a_{\vec k}^\dagger$, $\hat a_{\vec k}$ are the creation and annihilation operators of particles.
At zero temperature there is a true BEC in 2D, and we may use the standard Bogoliubov approach.
Assuming the weakly interacting regime where $mg/2\pi\hbar^2\ll 1$ and $r_*\ll \xi$, with $\xi=\hbar/\sqrt{mng}$ being the healing length. 
We may reduce the Hamiltonian (\ref{he3}) to a bilinear form, using the Bogoliubov transformation \cite{Huang}
$\hat a_{\vec k}= u_k \hat b_{\vec k}-v_k \hat b^\dagger_{-\vec k}-\beta_{\vec k}$,
where $\hat b^\dagger_{\vec k}$ and $\hat b_{\vec k}$ are operators of elementary excitations.
The Bogoliubov functions $ u_k,v_k$ are expressed in a standard way: $ u_k,v_k=(\sqrt{\varepsilon_k/E_k}\pm\sqrt{E_k/\varepsilon_k})/2$, 
$\beta_{\vec k}=\sqrt{n/S} U_k E_k/\varepsilon_k^2$, and  the Bogoluibov excitation energy is given by 
$\varepsilon_k=\sqrt{E_k^{2}+2ngE_k(1-Ck)}$. 

To zero order the chemical potential is $\mu=ng$.
If $C/\xi\leq \sqrt{8}/3$, $\varepsilon_k$ is a monotonic function of $k$.
However, it shows a roton-maxon structure for the constant $C$ in the interval $\sqrt{8}/3\leq C/\xi\leq 1$.
It is then convenient to represent $\varepsilon_k$ in the form \cite{boudjGora}:
\begin{equation}\label{spec} 
\varepsilon_k= \frac{\hbar^2 k}{2m}\sqrt{ (k-k_r)^2 +k_{\Delta}^2},
\end{equation}
where $k_r=2C/\xi^2$ and $k_{\Delta}=\sqrt{4/\xi^2-k_r^2}$.
If the roton is close to zero, then $k_r$ is the position of the roton, and 
$\Delta\!=\!\hbar^2 k_rk_{\Delta}/2m\!=\!2ngC\sqrt{mng/\hbar^2-C^2(mng/\hbar^2)^2}$, 
is the height of the roton minimum.
At $C/\xi=1$, the roton minimum touches zero and for $C/\xi >1$, the uniform Bose condensate becomes dynamically unstable 
and the uniform superfluid is no longer the ground state.\\
Importantaly, the spectrum (\ref {spec}) is independent of the random potential. 
This independence holds in fact only in zeroth order in perturbation theory; conversely, higher order calculations render the spectrum dependent on the random potential
due to the contribution of the anomalous terms $\langle\hat a_{-\vec k}\hat a_{\vec k}\rangle$.

The diagonal form of the Hamiltonian of the dirty dipolar Bose gas (\ref{he3}) can be written in the usual form 
$\hat H = E+\sum_{\vec k} \varepsilon_k\hat b^\dagger_{\vec k}\hat b_{\vec k}$,
where $E=E_0+\delta E+ E_R$ with $E_0= S g n^2/2 $ and $\delta E=\frac{1}{2}\sum\limits_{\vec k} [\varepsilon_k -E_k-ng(1-Ck)]$
being the ground-state energy correction due to qunatum fluctuations. 
\begin{equation}\label{Renergy} 
E_R=-\sum\limits_{\vec k} n\langle |U_k|^2\rangle \frac{ E_k}{\varepsilon_k^2} =-\sum\limits_{\vec k} n R_k \frac{ E_k}{\varepsilon_k^2},
\end{equation}
gives the correction to the ground-state energy due to the external random potential.

The noncondensed density is defined as  $\tilde{n}=\sum_{\vec k} \langle\hat a^\dagger_{\vec k}\hat a_{\vec k}\rangle$. 
Then invoking for the operators $a_{\vec k}$ the preceding Bogoliubov transformation, setting $\langle \hat b^\dagger_{\vec k}\hat b_{\vec k'}\rangle=\delta_{\vec k \vec k'}N_k$ 
and putting the rest of the expectation values equal to zero, where $N_k=[\exp(\varepsilon_k/T)-1]^{-1}$ are occupation numbers for the excitations.  
Using the fact that $2N (x)+1= \coth (x/2)$ \cite{boudj2015}, we obtain:
\begin{equation}\label {dep}
n'=\frac{1}{2S}\sum\limits_{\vec k} \left[\frac{E_k+ng(1-Ck)} {\varepsilon_k}\coth\left(\frac{\varepsilon_k}{2T}\right)-1\right] +n_R,
\end{equation}
where 
\begin{equation}\label {depdis}
n_R=\frac{1}{S}\sum\limits_{\vec k} \langle |\beta_k|^2\rangle=\frac{1}{S}\sum\limits_{\vec k} nR_k \frac{ E_k^2}{\varepsilon_k^4},
\end{equation}
is the condensate depletion due  to the external random potential.

%\section{fluctuations at zero temperature}

In order to investigate in a simple way how the random potential affects the behavior of the system, we
will often make the white-noise assumption in which the external potential is described by a single parameter $R (\vec r, \vec r')=R_0\delta (\vec r, \vec r')$ \cite{Gior}, 
where $R_0$ denotes the disorder strength which has dimension (energy) $^2$ $\times$  (length)$^2$.

Let us now asume that the roton is close to zero and the roton energy is $\Delta\ll ng$, we have the cofficient $C$  close to $\xi$, and $k_r\simeq 2/\xi$. 
Then, using Eq.(\ref{depdis}), for the contribution of momenta near the roton minimum at $T=0$, we obtain:
%and replacing the sum over $\vec k$ by the integral $\sum_{\vec k}=S\int d^2k/(2\pi)^2$, we obtain: 
\begin{equation} \label{depdis1}
\frac{n_R}{n}=\frac{mg}{4 \hbar^2}R \left(\frac{2ng}{\Delta}\right)^3,
\end{equation}
where $R=R_0/ng^2$ is a dimensionless disorder strength.\\
%We thus see that the density fluctuations grow logarithmically when the roton minimum is approaching zero and they can become strong for very small $\Delta$.
%In this case they lead to a significant depletion of the condensate. 
The noncondensed density is calculated via (\ref{dep}) as 
\begin{equation}    \label{dep1}
\frac{n'}{n} \simeq \frac{mg}{\pi\hbar^2}\left[\ln\left(\frac{2ng}{\Delta}\zeta\right)+\frac{\pi}{4} R\left(\frac{2ng}{\Delta}\right)^3\right];\,\,\Delta\ll ng,
\end{equation}
where $\zeta=\sqrt{2\pi\hbar^2/e^2 mg}$.\\
The leading term in Eq.(\ref{dep1}) which comes from the quantum fluctuations was first obtained in our recent work \cite {boudjGora}, 
while the second term which grows faster than the first one, represents the disorder correction to the condensate depletion.
The condensed fraction can be written as $n_c/n=1-n'/n$.

To check the result of Eq.(\ref{dep1}), we solve Eqs.(\ref {dep}) and (\ref{depdis}) numerically using Monte Carlo method. 
\begin{figure}[htb1] 
\includegraphics[scale=0.8]{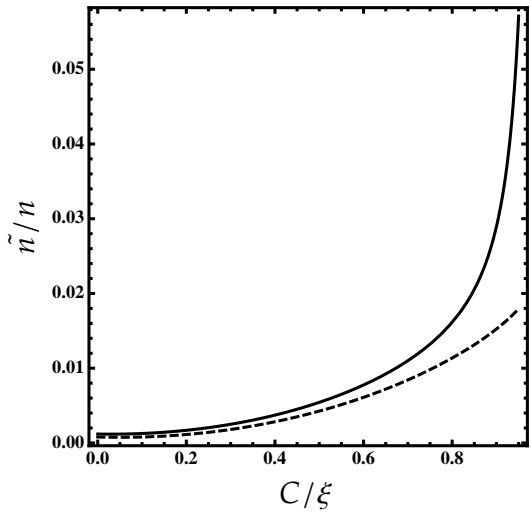}
\caption {Quantum depletion of the condensate from Eq. (\ref {dep}), as a function of $C/\xi$ for $mg/4\pi\hbar^2=0.01$.
Clean dipolar BEC:dashed line ($R=0$). Dirty dipolar BEC: solid line ($R=0.5$). }
\label{depl}
\end{figure}
Figure.\ref{depl} shows that  in the absence of the random external potential i.e. $R=0$, the noncondensed fraction grows logarithmically (see Eq. (\ref{dep1}))  
when the the roton energy $\Delta$ goes to zero yielding the transition to a supersolid state \cite {prok, boudjGora, petGora}.
%Our results are in good concordance with recent zero temperature phase diagram Monte Carlo simulations of \cite{bush, Astr}.  
It is worth stressing that in the context of the liquid helium, the position of the roton minimum influences also the phenomenon of superfluidity \cite{Noz}.\\
In the presence of the external disordered potential and for $C \approx \xi$, the condensate depletion becomes
more significant and diverges when $\Delta =0$ indicating the transition to a novel quantum phase in dilute 2D dipolar bosons. %(a superglass phase of dilute 2D dipolar bosons). 

%The integral over $dk$ is logarithmically divergent at large momenta because of the dipolar contribution to the interaction strength, $-gCk$. 
%However, this form of the dipole-dipole contribution is valid only for $k\ll 1/r_*$. We thus may put a high momentum cut-off $1/r_*$, which leads for $\Delta\ll ng$:

%Thus, the inequality (\ref{ineq1}) indicates that at zero temperature only for $\Delta$ extremely close to zero the fluctuations are so strong that the Bogoliubov approach fails (see Fig.3).

\begin{figure}[htb1]
\centering{
\includegraphics[scale=0.8]{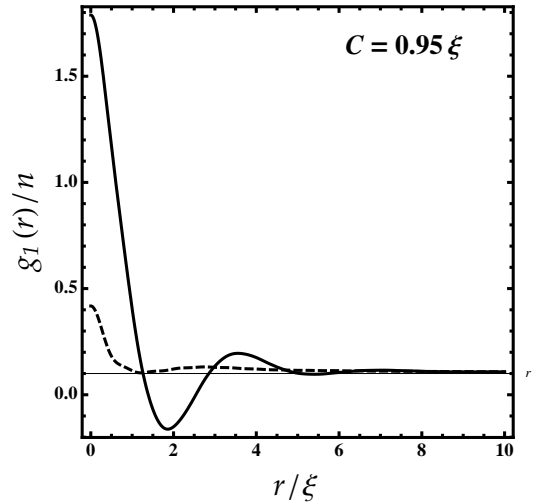}}
\caption {One-body density correlation function for $mg/4\pi\hbar^2=0.01$. 
Clean dipolar BEC:dashed line ($R=0$). Dirty dipolar BEC: solid line ($R=0.5$).}
\label{gd}
\end{figure}

The same factors of Eq.(\ref{dep1}) appear in the one-body density matrix $g_1(\vec r)=\langle\hat\Psi^{\dagger}({\bf r})\hat\Psi(0)\rangle$,
where $\hat\Psi({\bf r})$ is the field operator. The correction due to the disorder effects to the correlation function is  $g_1^R (\vec r)=\int n_R e^{i \vec k \vec r} (d^2k/ (2\pi)^2$.
 Assuming that the roton minimum reaches zero and taking into account only the contribution of momenta near this minimum we have for $\Delta\ll ng$:
\begin{equation}   \label{g1}
\frac{g_1(r)} {n}=1+\frac{mg}{\pi\hbar^2}\left[\ln\left(\frac{2ng}{\Delta}\right)+\frac{\pi}{4} R\left(\frac{2ng}{\Delta}\right)^3\right] J_0(2r/\xi),
\end{equation}
where $J_0(x)$ is the Bessel function.\\
The numerical calculation of the one-body density matrix well agrees with the analytical result obtained in Eq (\ref{g1}) 
and shows that $g_1(r)$ decays at large distance when $C\approx \xi$ as is depicted in Fig.\ref {gd}. 
This signals the non-existence of the off-diagonal long-range order (i.e., BEC) in the disordered 2D dipolar bosons.            

As we see from Eq.~(\ref{dep1}), for the roton minimum close to zero a small condensate depletion requires the inequality 
\begin{equation}   \label{Bogcrit0}
%\frac{mg}{\hbar^2}\ln\left(\frac{2ng}{\Delta}\zeta\right) \ll 1, \,\,\,\,\,\, 
\frac{mg}{\hbar^2} R\left(\frac{2ng}{\Delta}\right)^3 \ll 1.
\end{equation}
We thus conclude that at $T=0$, the validity of the Bogoliubov approach is guaranteed by the presence of the small parameter (\ref{Bogcrit0}).

However, the situation changes in the calculation of the correction to the ground-state energy due to the external random potential.
When the roton minimum is approaching to zero, we get from (\ref{Renergy})
\begin{equation}   \label{Renergy1}
\frac{E_R}{E_0}=-\frac{2mg}{\hbar^2} R \left(\frac{2ng}{\Delta}\right);\,\,\,\,\,\,\Delta\ll ng.
\end{equation}
Equation (\ref {Renergy1}) shows that $E_R$ grows linearly with $ng/\Delta$, and has a negative value which leads to reduce the total energy of the system.\\
The correction to the ground-state energy due to quantum fluctuations can be given as 
\begin{equation}   \label{energy1}
\frac{\delta E}{E_0}\simeq 1+\frac{2mg}{\pi\hbar^2}+\frac{2mg}{\pi\hbar^2}\ln\left(\frac{2ng}{\Delta}\right);\,\,\,\,\,\,\Delta\ll ng.
\end{equation}
The disorder correction to the chemical potential can be calculated esealy  through $\partial E_R/\partial N$ 
\begin{equation}    \label{deltamu}
\frac{\mu_R}{\mu} \simeq -\frac{mg}{\hbar^2} R\left(\frac{2ng}{\Delta}\right)^3=-4n_Rg;\,\,\,\,\,\,\Delta\ll ng.
\end{equation} 
Note that quantum fluctuations corrections to the chemical potential had already obtained in our recent paper \cite{boudjGora}.

One can also show that the shift of the sound velocity is consistent with the change in the compressibility $mc_s^2 = n \partial \mu/\partial n $ \cite {boudj2015,Lev} and is given by
\begin{equation}    \label{sound}
\frac{c_s^2}{c_{s0}^2} = 1+\frac{mg}{\pi\hbar^2}\left[2\ln\left(\frac{2ng}{\Delta}\right)+\left(\frac{2ng}{\Delta}\right)^2-\frac{3\pi}{2} R\left(\frac{2ng}{\Delta}\right)^5\right],
\end{equation} 
where $c_{s0}=\sqrt{\mu/m}$ is the zeroth order sound velocity. 
The second and the third terms originate from quantum fluctuations while the last term comes from the disorder contribution.

%\section{Fluctuations at finite temperatures}

On the other hand,  in an infinite uniform 2D fluid thermal fluctuations at any nonzero temperature are strong enough
to destroy the fully ordered state associated with BEC, but are not
strong enough to suppress superfluidity in an interacting system at low, but non-zero
temperatures. The presence of residual "quasi-long-range" order at low temperatures
leads to an interesting interplay between superfluidity and condensation in all experimentally
relevant finite-size systems. In this quasicondensate, the phase coherence governs only regimes of a size smaller than
the size of the condensate, characterized by the coherence length $l_\phi$\cite {petr1, boudj2012}.
Thermodynamic properties, excitations, and correlation properties on a distance scale smaller than $l_{\phi}$ are the same as in the case of a true BEC. 
%Moreover, for realistic parameters of quantum gases, $l_{\phi}$ exceeds the size of the system \cite{GPS}, so that one can employ the ordinary BEC theory. 
Upon utilizing the previous definitions we find that the correction to the condensate depletion, the correlation function and thermodynamic quantities due to thermal flucatuations 
is given by the factor $(2mg/\hbar^2)T/\Delta$ \cite{boudjGora}.

The superfluid fraction $n_s/n$ can be found from the normal fraction $n_n/n$ which is determined by the transverse current-current
correlator $n_s/n =1-n_n/n$. We apply a Galilean boost with the total momentum of the moving system ${\bf \hat P_v}={\bf \hat P}+mv N$, where 
${\bf \hat P}=\sum_k {\bf k} \hat a^\dagger_{\vec k} \hat a_{\vec k}$ and $v$ is the liquid velocity. In $d$-dimensional case, the superfluid fraction reads

\begin{equation}   \label{sup}
 \begin{split}
 \frac{n_s}{n}= 1-\frac{2}{dTn} \int \frac{d^dk}{(2\pi)^d} \left[\frac{E_k}{4 \text {sinh}^2 (\varepsilon_k/2T)} \right. \\+
\left. \frac{n R_0 E_k^2}{\varepsilon_k^3}  \text {coth} \left(\frac{\varepsilon_k}{2T}\right) \right].
\end{split}
\end{equation}
At very low temperature we can put $\text {coth}(\varepsilon_k/2T)=2T/\varepsilon_k$. Thus, Eq. (\ref{sup}) reduces to   
\begin{equation}  \label{sup1}
 \frac{n_s}{n}= 1-\frac{4}{d}\frac{n_R}{n} -\frac{2}{dTn}\int \frac{d^dk}{(2\pi)^d} \left[ \frac{E_k}{4 \text {sinh}^2 (\varepsilon_k/2T)}\right]. 
\end{equation}
Interestingly, the ratio between the normal fluid density and the corresponding condensate depletion increases to 2 in 2D and to 4 in 1D, in contrast to the familiar 4/3
in 3D geometry obtained earlier in \cite {Huang, Gior}. Another important remark is that the superfluid fraction (\ref{sup1}) is no longer a tensorial quantity 
as in the case of a 3D dirty dipolar Bose gas \cite{AxelM, Boudj} since the dipoles are assumed to be perpendicular to the plane. 
However, if the dipoles would be tilted slightly, superfluidity would acquire an anisotropy and thus becomes a tensorial quantity.

Assuming now that the roton minimum is close to zero and $\Delta\ll T$, then the momenta near the roton minimum are the most important, and the use of Eq.(\ref{depdis1}) yields:
\begin{equation}   \label{super} 
\frac{n_s}{n}= 1-\frac{mg}{2 \hbar^2}R \left(\frac{2ng}{\Delta}\right)^3-\frac{2mg}{\hbar^2}\frac{T}{\Delta}. 
\end{equation}
At $T=0$ and in the absence of disorder, the superfluid fraction is equal to unity. 
Another interesting consequence of the above results is that for any value of $(ng/\Delta)$ there exists
a critical strength of disorder $R_c=(4/\pi) \ln (2ng\zeta/\Delta)/ (2ng/\Delta)^3$ for which $n_s/n < n_c/n$.
For dysprosium atoms at 2D densities $\sim 10^9$ cm$^{-2}$, one has  $r_*\simeq 200$ \AA, $ng \sim$5 nK, $mg/2\pi\hbar^2\simeq 0.025$
and $\Delta$ should be above 2 nK. Therefore, the precedent criterion is satisfied for $R_c= 8\times10^{-4}$, which means that $n_s$ can be smaller than $n_c$ 
only for a very weak disorder i.e. $R_c \ll 1$.\\
One can conclude that the  Bogoliubov approach should satisfy the conditions $R<R_c$.
However, it is not clear whether these results still apply for $R>R_c$ in a range of densities where the difference between
$n_s/n$ and $n_c/n$ can be significant. The response to these questions requires either a non-perturbative scheme or numerical Quantum Monte Carlo simulations, 
which are out of the scope of the present work.

%\begin{figure}[htb1] 
%\includegraphics[scale=0.8]{superfluid.eps}
%\caption {Superfluid fraction from Eq.(\ref {sup}), as a function of $C/\xi$ and for $T/ng$=0.1. Dashed line: $R'$=0.5. Solid line: $R'$=3. }
%\label{supg}
%\end{figure} 

%Figure.\ref{supg} shows that at low-temperature and for small value of disorder strength ($R'$=0.5), the system is purely superfluid where $n_s/n$ remains almost constant. 
%For $R'$=3, the system is characterized by the existence  of two phases at the same time: a superfluid phase for $C \leq 0.9\xi$ where $n_s/n$ decreases progressively. 
%On the other hand, when the roton minimum is very close to zero $C \sim \xi$, the superfluid fraction vanishes which means that the condensed particles are localized and form the
 %so-called Bose glass state. This clearly indicates the occurrence of the SG state.

%\section{Conclusions}

In conclusion, the impact of a weak random potential on BEC and on superfluidity in a dilute quasi-2D dipolar Bose gas is studied with a combined numerical and analytical approach.
Our analysis signifies a more pronounced effect of disorder in such a system when the roton is approching zero 
with enhancing quantum fluctuations, one-body density-matrix, equation of state, sound velocity and depleting superfluid density. 
Furthermore, we have reproduced the expression of the condensate fluctuations and thermodynamic quantities obtained 
in our recent work \cite {boudjGora} in the absence of the disorder potential.
Our results represent a starting point for the analysis of collective modes of homogeneous or trapped 2D dirty dipolar BECs.
We have found, in addition,  the criterion of applicability of the Bogoliubov approach for these systems.
An important step for a future work is to show whether these disordered quasi-2D dipolar systems promote a stable superglass phase, where superflow and glassy
density localization coexist in equilibrium without exhibiting phase separation. A qualitative study of such a phase diagram  
necessitates a more sofisticated scheme with a non-perturbative solution. 
%Promising candidates for the creation of such SG state is the dysprosium atoms which has a small dipole-dipole distance $r_*\simeq 200$ \AA. 
%Hence, at 2D densities $\sim 10^9$ cm$^{-2}$ one has $nr_*\ll 1$ and the roton-maxon spectrum is realized for the 3D scattering length $a_{3D}$ 
%of several tens of angstroms at the frequency $\omega=10$ kHz leading to the confinement length $l_0$ about 1000 \AA. 

%\section*{Acknowledgements}
We are grateful to Nikolay Prokofiev and Axel Pelster for interesting discussions and for comments on the manuscript.

\end{document}